\documentclass[useAMS,usenatbib]{mn2e}
\pdfoutput=1                          

\usepackage{graphicx}
\usepackage{aasmacros}
\usepackage{color}

\def\hmpc{$h^{-1}$Mpc}

\def\msol{M$_\odot$}
\def\hmsol{$h^{-1}$M$_\odot$}

\def\nsat{\langle N_{\rm sat}\rangle_M}

\def\om{\Omega_m}

\def\s8{\sigma_8}

\def\lcdm{$\Lambda$CDM}

\def\x2{$\chi^2$}
\def\hmsol{$h^{-1}\,$M$_\odot$}

\def\NNm1{\langle N(N-1) \rangle}

\def\m_star{M_\ast}

\def\lcdm{$\Lambda$CDM}

\def\om{\Omega_m}

\def\s8{\sigma_8}

\def\hmpc{$h^{-1}\,$Mpc}

\def\x2{$\chi^2$}
\def\hmsol{$h^{-1}\,$M$_\odot$}

\def\nsat{\langle N_{\mbox{\scriptsize sat}}\rangle_M}

\def\NNm1{\langle N(N-1) \rangle}

\def\p0{P_0(r)}

\def\fq{f_{\rm Q}}
\def\fqcen{f_{\rm Q}^{\rm cen}}
\def\mgal{M_\ast}

\def\msol{{\rm M}_\odot}
\def\dn{{\rm D}_n4000}

\def\mhalo{M_{h}}
\def\lesssim{\la}

\def\psat{P_{\rm sat}}
\def\fqb{f_{\rm Q}^{\rm blue}}
\def\fqr{f_{\rm Q}^{\rm red}}
\def\zhalf{z_{1/2}}
\def\mpeak{M_{\rm peak}}
\def\nsat{N_{\rm sat}}
\def\ssfr_fib{sSFR$^{\rm(fib)}$}
\def\mpc{\rm Mpc}

\title[Galactic Conformity and Assembly Bias]{Halo Histories vs. Galaxy Properties
  at z=0\\ II: Large-Scale Galactic Conformity}

\author[Tinker et.~al.]{\parbox{\textwidth}{Jeremy L. Tinker$^1$, ChangHoon Hahn$^1$, Yao-Yuan
  Mao$^{2}$, \\ Andrew R. Wetzel$^{3,4,5}$\thanks{Caltech-Carnegie Fellow}, Charlie Conroy$^6$\\
\footnotesize
$^1$Center for Cosmology and Particle Physics, Department of Physics, New York University, New York, NY\\
$^2$Department of Physics and Astronomy \& Pittsburgh Particle Physics, Astrophysics, and Cosmology Center (PITT PACC), University of Pittsburgh, PA 15260
$^3$Carnegie Observatories, Pasadena CA\\
$^4$TAPIR, California Institute of Technology, Pasadena, CA\\ 
$^5$Department of Physics, University of California, Davis, CA\\
$^6$Department of Astronomy, Harvard University, Cambridge, MA
}}

\begin{document}


\pagerange{\pageref{firstpage}--\pageref{lastpage}} \pubyear{2016}

\maketitle

\label{firstpage}

\begin{abstract}

  Using group catalogs from the SDSS DR7, we attempt to measure
  galactic conformity in the local universe. We measure the quenched
  fraction of neighbor galaxies around isolated primary galaxies,
  dividing the isolated sample into star-forming and quiescent
  objects. We restrict our measurements to scales $>1$ Mpc to probe
  the correlations between the formation histories of distinct
  halos. Over the stellar mass range $10^{9.7} \le \mgal/\msol \le
  10^{10.9}$, we find minimal statistical evidence for conformity. We
  further compare these data to predictions of the halo age-matching
  model, in which the oldest galaxies are associated with the oldest
  halos at fixed $\mgal$. For models with strong correlations between
  halo and stellar age, the conformity signal is too large to be
  consistent with the data. For weaker implementations of
  age-matching, galactic conformity is not a sensitive diagnostic of
  halo assembly bias, and would not produce a detectable signal in
  SDSS data. We reproduce the results of \cite{kauffmann_etal:13}, in
  which the star formation rates of neighbor galaxies are
  significantly reduced around primary galaxies when the primaries are
  themselves low star formers. However, we find this result is mainly
  driven by contamination in the isolation criterion; when using our
  group catalog to remove the small fraction of satellite galaxies in
  the sample, the conformity signal largely goes away. Lastly, we show
  that small conformity signals, i.e., 2-5\% differences in the
  quenched fractions of neighbor galaxies, can be produced by
  mechanisms other than halo assembly bias. For example, if passive
  galaxies occupy more massive halos than star forming galaxies of the
  same stellar mass, a conformity signal that is consistent with
  recent measurements from PRIMUS (\citealt{berti_etal:16}) can be
  produced.

\end{abstract}


\section{Introduction}

Galaxy evolution is indelibly linked to the evolution of the dark
matter structure in which they form. The purpose of this series of papers is to
quantify the degree of correlation between galaxy properties and halo
properties in the local universe, and through this investigation make
inferences about the correlated evolutionary histories of both. The
key tool that we use in this series is a galaxy group finder, which,
when applied to a statistical sample of galaxies, can robustly
determine which galaxies are central, meaning they exist at the center
of a distinct dark matter halo, and those galaxies that are satellites, meaning
they orbit within a larger dark matter halo. 

In Paper I of this series (\citealt{tinker_etal:16_p1}), we measured
the quenched fraction of central galaxies as a function of large-scale
environment. The correlation between large-scale density and galaxy
properties is well known: galaxies in denser environments are
preferentially quenched of their star formation and elliptical in
their morphology (see, e.g., \citealt{blanton_moustakas:09} and
references therein). However, when broken down into the relative
contribution of central and satellite galaxies, the quenched fraction
of central galaxies is nearly independent of environment. The observed
correlations can be explained by the increasing fraction of satellite
galaxies at high densities, which are preferentially quenched
(\citealt{tinker_etal:11}). We compared these measurements to models
in which halo age is matched to galaxy age at fixed galaxy and halo
masses: the older halos contain galaxies the quenched galaxies, while
the most rapidly growing halos contain the most actively star-forming
galaxies. This is known as the `age-matching' model
(\citealt{hearin_watson:13}). The interest in the age-matching model
centers on the fact that the model makes testable predictions for the
spatial clustering of active and passive galaxies: at fixed mass,
older halos are more strongly clustered than their younger
counterparts, an affect known as assembly bias (see, e.g.,
\citealt{wechsler_etal:06, gao_white:07, li_etal:08}). Thus, the
galaxies that occupy these halos--the central galaxies---would show a
clear correlation between their quenched fraction and their
large-scale density, with most quenched central galaxies being in
high-density regions. In the age-matching model, this correlation is
expected to be strongest for lower mass galaxies, where assembly bias
in dark matter halos is strongest.  In Paper I, we found the
observations were not consistent with predictions of the age-matching
model at $\mgal\la 10^{10.3}$ $\msol$. We compared these measurements
to a wide variety of halo age definitions. At higher galaxy masses, there
was a weak trend of $\fqcen$ with $\rho$, a correlation consistent
with an age-matching model in which halo age was defined in such a way
as to minimize assembly bias within the halo population. The
implication of these results is that the mechanism that quenches
galaxies is uncorrelated with halo formation history at low masses,
and only weakly correlated at higher masses.

In this paper we probe a complementary observable for detecting
assembly bias within the galaxy population: galactic conformity.
Galactic conformity is the observed correlation between the properties
of separate galaxies. Using group catalogs, \cite{weinmann_etal:06a}
noted that the colors of satellite galaxies within the group were more
likely to `conform' to the color of the central galaxy at fixed halo
mass. These measurements have been confirmed by a number of other
studies (\citealt{knobel_etal:15, kawinwanichakij_etal:16,
  berti_etal:16}) \cite{kauffmann_etal:13} (hereafter K13) measured conformity between
the star formation rates of galaxies separated by up to 5 \mpc---well
outside the virial radius of the primary galaxy's halo. This
large-scale conformity has been proposed as a test of galaxy assembly
bias (\citealt{hearin_etal:15}). The results of Paper I indicate that
galaxy quenching is, at most, weakly correlated with large-scale
environment and, by extension, halo formation history. In this paper,
we will present new measurements of galactic conformity, as well as a
critical examination o the K13 result. Additionally, we will explore
sources of a `conformity signal' that do not arise from assembly bias.


In this work, as in Paper I, we focus on samples of central galaxies
in narrow bins of stellar mass. Because the goal is to determine whether halo assembly
bias has an impact on galaxy formation, defining the problem in this manner minimizes
possible systematic biases in the measurements. We will test for
conformity in two different properties of galaxies: galaxy quenched
fraction (as in Paper I) and galaxy specific star formation rate (as
used in \citealt{kauffmann_etal:13}). In the latter, we focus on
reproducing the K13 measurement and explaining the result in the
context of our group catalogs.

\begin{figure*}
\includegraphics[width=7in]{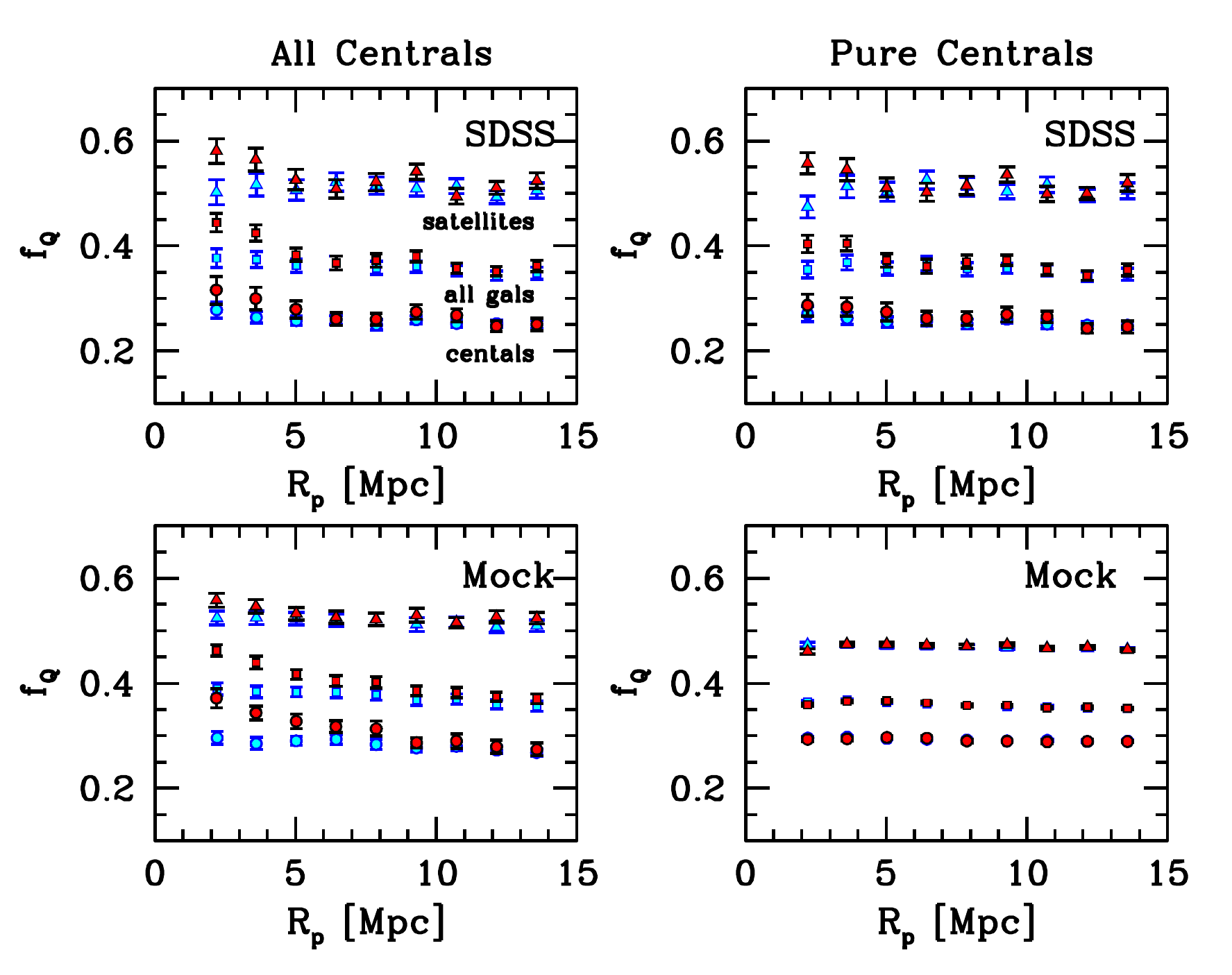}
\caption{ \label{conformity_tests} The conformity signal around both
  mock galaxy catalogs and SDSS galaxies of mass $\mgal=10^{10}$
  $\msol$, in a bin 0.2 dex wide. We show $\fq$ of secondary galaxies
  around primary central galaxies as a function of projected
  separation. In each panel, from top to bottom, the secondary
  galaxies are satellite galaxies (triangles), all galaxies (squares),
  and central galaxies (circles). Assembly bias, if present, should
  primarily effect $\fq$ for central secondaries. Blue symbols
  represent $\fq$ around primaries that are star-forming, while red
  symbols represent primaries that are quenched. The left-hand panels
  show the results when the sample of primary galaxies includes all
  central galaxies of $10^{10}$ $\msol$, while the right-hand panels
  show the results when the primary sample is restricted to `pure'
  central galaxies, which are centrals with a $\psat<0.01$. The mock
  galaxy catalog has no assembly bias in it, thus the conformity
  `signal' seen on the left-hand side is purely an artifact of
  impurities in the group finder. This bias is gone when restricting
  the primaries to pure centrals. }
\end{figure*}

\section{Data, Measurements, and Methods}

\subsection{Galaxy Groups from DR7}

To construct our galaxy samples, we use the NYU Value-Added Galaxy
Catalog (VAGC; \citealt{blanton_etal:05_vagc}) based on the
spectroscopic sample in Data Release 7 (DR7) of the Sloan Digital Sky
Survey (SDSS; \citealt{dr7}).  The details of these catalogs, and the
algorithm for finding the groups, can be found in
\cite{tinker_etal:11}, \cite{campbell_etal:15}, and
Paper I. In brief, we create volume-limited samples
that are complete in stellar mass, within which the groups are
identified. The group finding algorithm assigns probabilities to each
galaxy quantifying the likelihood that a galaxy is a satellite, $\psat$. To create the full central-satellite breakdown
of the entire galaxy population, galaxies with $\psat<0.5$ are
classified as central. However, at this threshold for central
classification, there are impurities in the sample. To attenuate this
effect, we will use galaxies with $\psat<0.01$. We will refer to these
objects as `pure centrals'. This restriction yields only a modest
reduction of the number of centrals in the sample; $\sim 90\%$ of
centrals in the sample are pure. This extra restriction removes
central galaxies from the primary sample that are within the projected
radius of a larger group but separated in $\Delta v$ by values larger
than one to two times the velocity dispersion of the larger halo,
depending on how close to the radial edge of the larger halo it
lies. This is primary source of impurities in the central galaxy
sample, and restricting our sample to pure centrals increases the
purity of the sample to $\sim 99\%$.

We use stellar masses from the NYU VAGC, which are in turn created by
the code of \cite{blanton_roweis:07}. We use $\dn$ as our proxy for
identifying galaxies quenched of their star formation. $\dn$, taken
from the MPA-JHU SDSS spectral reductions\footnote{\tt
  http://www.mpa-garching.mpg.de/SDSS/DR7/}
(\citealt{brinchmann_etal:04}), is a more robust indicator of galaxy
quiescence because it is less susceptible to dust contamination than
broadband colors (e.g., \citealt{maller_etal:09,
  masters_etal:10_dust}). We define a galaxy as quenched if $\dn>1.6$,
a value robustly marks the minimum between the bimodal distribution
between the red sequence and the star-forming main sequence.

\begin{figure*}
\vspace{-2.5cm}
\includegraphics[width=6.0in]{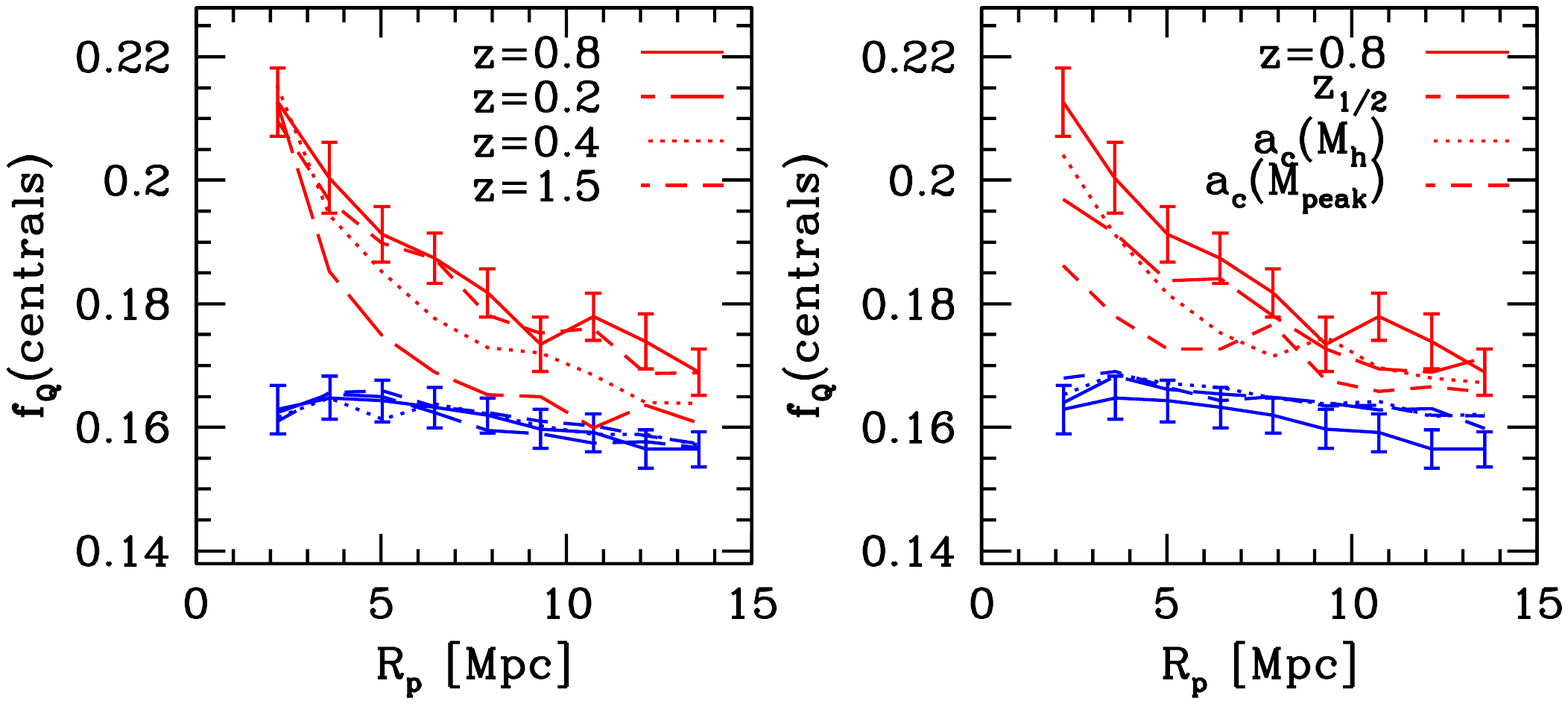}
\vspace{-10.cm}
\caption{ \label{conformity_age} The conformity signal in a bin of
  $\log\mgal/\msol=[9.7-9.9]$ for different implementations of the
  age-matching model. All results are obtained after running the group
  finder on each mock and analyzing the resulting mock group
  catalog. All results use a primary sample of pure centrals and a
  secondary sample of all centrals. {\it Left Panel:} Results of
  age-matching models where halo age is defined as fractional growth
  over a redshift baseline. Halos with the least amount of growth over
  this timeframe are ranked as the oldest. {\it Right Panel:} Results
  of age-matching models where age is defined as the half-mass
  redshift, $\zhalf$, (long-dash), the formation epoch of
  \citet{wechsler_etal:02}, $a_c(\mhalo)$ (dotted), and the formation
  epoch using $\mpeak(z)$ (short-dashed) rather than the current mass,
  $\mhalo$. $\mpeak$ is always monotonically increasing by
  construction. }
\end{figure*}

\subsection{Mock Galaxy Samples and the Age-Matching Model}

In this paper we will compare the results from the group catalog to
expectations from dark matter halos. For most results, we use the
`Chinchilla' simulation, 
the same simulation used in Paper I. The box size is 400 \hmpc\ per side,
evolving a density field resolved with $2048^3$ particles, yielding a
mass resolution of $5.91\times 10^8$ \hmsol. The cosmology of the
simulation is flat \lcdm, with $\om=0.286$, $\sigma_8=0.82$, $h=0.7$,
and $n_s=0.96$.  As in Paper I, halos are found in the simulation
using the Rockstar code of \cite{rockstar} and Consistent Trees
(\citealt{consistent_trees}) is used to track halo growth. In \S 3.3 we
will use the the MultiDark Planck simulation
(\citealt{klypin_etal:16}) to create mock galaxy samples at $z\sim
0.3$. We will discuss the pertinent details of that simulation and
its usage in that section.

In Paper I we compared measurements directly to statistics of
halos. Here, we create full mock galaxy catalogs that are processed
through the group finding algorithm to incorporate and test any
observational biases that arise in this procedure. To assign central
galaxies to each halo, we do the following: First, we use the results
of the observed group catalog to determine the relationship between
host halo mass and central galaxy stellar mass for $\mgal\ge 10^{9.7}$
$\msol$.  Halos in the simulation are matched to halos of the same
mass in the group catalog, thus any scatter found in the group catalog
is preserved in the mock. Once the stellar masses of the central
galaxies have been assigned, the mock central galaxies and the group
catalog central galaxies are divided into bins of 0.1 dex of stellar
mass.  In each bin, the mock central galaxies are rank-ordered by the
age of their halos (which we will define below). Once ranked, values
of $\dn$ are assigned to the mock central galaxies by matching the
rank-ordered lists of halo age to group catalog $\dn$: the oldest halo
is assigned the highest value of $\dn$, and on down the list. This
method is consistent with the age-matching model of
\cite{hearin_watson:13} and yields a conformity signal similar to
those presented in \cite{hearin_etal:15}. We also have a mock with no
assembly bias, in which $\dn$ values are assigned randomly in each bin
of stellar mass. In this latter model, there should be no conformity
signal because the probability of being quenching is uncorrelated with
the halo age.

To assign satellites to each halo, we first measure the mean number of
satellites at $\mgal\ge 10^{9.7}$ $\msol$ as a function of halo mass
in the SDSS group catalog. For each halo in the simulation, we
randomly draw a Poisson deviate around the mean to represent the
number of satellites in that halo, $\nsat$. From the group catalog, we bin all
satellites by their host halo mass. For each simulated halo, we
randomly draw actual satellites from the halo mass bin corresponding
to that halo, up to $\nsat$. Thus, each satellite in the simulated
halos has the values of $\mgal$ and $\dn$ from the SDSS
satellite. There is no assembly bias in the satellite galaxies---i.e., whether a satellite is star forming or quenched is independent of the properties of the host halo. As
shown in \cite{hearin_etal:15}, satellite assembly bias has
minimal---if any---impact on large-scale galactic conformity. 
 
We define halo age using various definitions, all of which are
discussed in detail in Paper I. These age definitions fall into two
distinct classes: (1) halo growth over a redshift baseline, and (2)
proxies for `formation epoch' of the halos. For (1), our fiducial
model rank-orders halos by their growth since $z=0.8$. We also
investigate other baselines using $z=0.2$, 0.4, and 1.5. A redshift
baseline of $z=0.8\rightarrow 0$ roughly spans the range over which
most $\mgal\lesssim 10^{11.3}$ $\msol$ central galaxies arrive on the
red sequence (\citealt{tinker_etal:13_shmr}). Timescales from lower
redshifts reflect short-term growth, near timescales estimated for the
quenching timescale of galaxies (\citealt{peng_etal:15,
  hahn_etal:16_tq}). Longer baselines are closer to the half-mass
redshifts of $\mhalo\sim 10^{12}$ $\msol$ halos. For (2), we use the
half-mass redshift itself, $\zhalf$, which is the most common age
definition in the literature. We also use two different versions of
the formation epoch defined in \cite{wechsler_etal:02}, $a_c$, which
we describe presently.

The typical implementation of $\zhalf$ or $a_c$ uses the redshift
evolution of the halo itself, $\mhalo(z)$. However, this quantity is
not always monotonically increasing. Tidal encounters with larger
halos, or even `splashback' events, in which halos actually pass
through a larger halo and emerge back out, can strip mass off the
halo. Thus, $\mhalo$ at $z=0$ may be lower than the peak halo mass,
$\mpeak$. $\mpeak(z)$ is defined as the highest halo mass at any time
$\ge z$, and it is a monotonically increasing function of time. Halos
that have encountered significant stripping will be ranked very high
when identifying the `oldest' objects. In Paper I we showed that this
is what drives the very strong assembly bias signal in low-mass
halos. Using halo growth histories as a function of $\mpeak(z)$ rather
than $\mhalo(z)$ removes the effects of these types of encounters. In
\cite{wetzel_etal:14}, we demonstrated that splashback encounters have
little immediate impact on the galaxy star formation rate; such
objects quench the same as 'normal' satellites, in that there is a
delay of several Gyr before any quenching of star formation
begins. Thus, in Paper I we concluded that a more physically realistic
age-matching model should use $\mpeak(z)$ rather than $\mhalo(z)$ to
determine halo age. The $\mpeak$ age-matching model was in good
agreement with measurements of the dependence of $\fq$ on large-scale
density for high-mass galaxies ($\mgal \ga 10^{10.3}$ $\msol$, where
the $\fq$ for central galaxies goes above 50\%). However, even
$a_c(\mpeak)$ produced a correlation between $\fq$ and $\rho$ much stronger than that measured in 
galaxies at lower stellar masses.

After populating the halos with mock galaxies, the galaxies are
projected into an angular space, giving each galaxy an RA, Dec, and
$z$, covering a total area of $5,156$ deg$^2$ (1/8 of the sky)
and extending to a maximum redshift of $z=0.138$, which corresponds to
a comoving radius of the box length, 400 \hmpc. Each mock is then
passed through the group finder. All conformity measurements from the
mocks are measured on the mock group catalogs in order to incorporate
any biases imparted by the group-finding process (see
\citealt{campbell_etal:15} for a thorough assessment of the precision
and accuracy of the group finder used here as well as other finders in
the literature). Additionally, the group finding process does not use
$\dn$ information at all, only positions, velocities, and stellar
masses. Thus, even if the assignment of $\dn$ values is biased, this
does not impact the resulting group catalog.

\subsection{Making Conformity Measurements}

To quantify conformity, we measure the fraction of `secondary'
galaxies that are quenched, $\fq$, as a function of projected
separation, $R_p$, from `primary' galaxies. Primary galaxies are
divided into quenched or star-forming samples. We will refer to $\fq$
around each type of primary galaxy as $\fqr$ and $\fqb$,
respectively. To be clear, primary and secondary samples do not imply
mutually exclusive sets of galaxies. In our measurements, secondary
galaxies are defined as central galaxies of the same mass range as the
primary galaxies.  Primary galaxies can be a secondary to another
primary. We will show presently that our choices of primary
and secondary galaxies remove observational biases.

At each bin in $R_p$, we include galaxies with $\Delta v\le 500$ km/s
with respect to the primary galaxy.  We make all measurements in bins
of fixed stellar mass. Our goal in this paper is to use conformity as
a test of assembly bias. Halo assembly bias is the effect that the
clustering of galaxies at fixed halo mass depends on halo formation
history. Restricting the galaxy sample to only centrals brings the
sample closer to a sample of host halos. Performing the measurements
in bins of fixed stellar mass is a rough approximation for fixing halo
mass.

Figure \ref{conformity_tests} shows and example of our conformity
measurements for $\log\mgal/\msol=[9.7,9.9]$ for both mock galaxies
and the SDSS group catalog (we will discuss the SDSS measurements in
detail in the Results section, \S 3). The top two panels show the
results measured from the SDSS group catalogs while the bottom two
panels show results from one our mock galaxy catalogs. The left-hand
panels show results when the primary sample is made up of all central
galaxies, and the right-hand panels show results when the primary
sample is restricted to pure centrals. In each panel, we show $\fq$
where the secondaries are centrals (bottom), satellites (top), and all
galaxies (middle).

We first discuss the mock results. The mock catalog used in Figure
\ref{conformity_tests} contains no assembly bias; i.e., $\dn$ is
uncorrelated with any halo age proxy. Thus, the mock contains no
intrinsic galactic conformity, and any difference in $\fqr$ and $\fqb$
is due entirely to biases from the group finder. Thus, the rise in
$\fqr$ for all centrals is due to misclassification of centrals and
satellites within the primary galaxy sample. This is an example of
observational biases that can result from how primary galaxies are
identified. However, when restricting the primaries to pure centrals,
all biases are eliminated. The results for $\fqr$ and $\fqb$ are
independent of $R_p$ and consistent with each other.

The results for the SDSS group catalog are quantitatively similar to
the mock results. When using all centrals as the primary sample, the
quenched fraction around red primaries rises as $R_p$ approached the
virial radius of the halo, diverging from the quenched fraction around
star forming primaries. These trends are largely removed when shifting
the primary sample to pure centrals. We will discuss these results
more quantitatively in \S 4.

\begin{figure*}
\vspace{-3cm}
\includegraphics[width=7in]{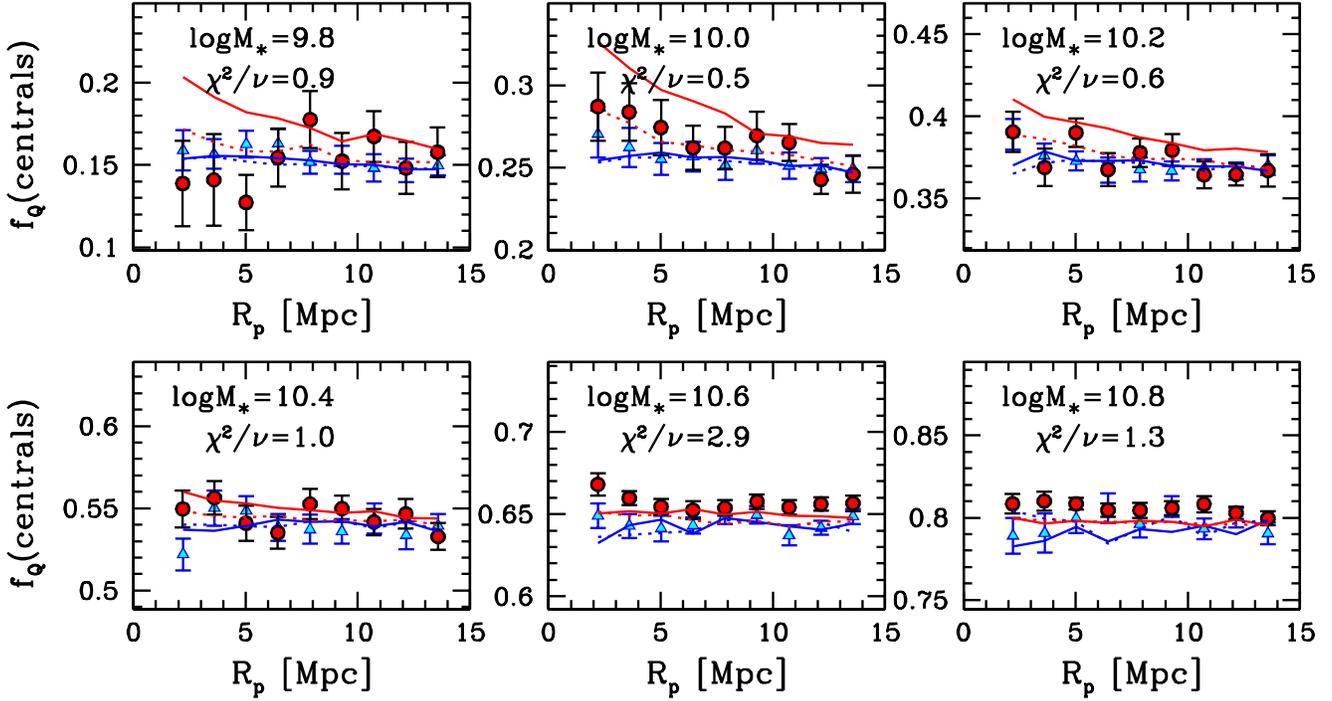}
\vspace{-9.2cm}
\caption{ \label{conformity_data} Conformity signal around primary
  galaxies as a function of stellar mass. In each panel, the primaries
  are pure centrals, while secondaries are all centrals of the same
  stellar mass. The symbols represent measurements from the SDSS group
  catalog, while solid curves show predictions from the age-matching
  model using $\Delta z=0.8$ as the age proxy. Dotted curves show the
  age-matching predictions for $a_c(\mpeak)$. Red symbols and curves
  show $\fq$ around quenched primaries, while blue symbols and curves
  show $\fq$ around star forming primaries. Error bars are obtained by
  jackknife sampling. In each panel, the value of $\chi^2/\nu$ is
  obtained by comparing $\fqb$ and $\fqr$. Values over unity indicate
  a statistically significant difference between the two quantities,
  and thus imply a detection of conformity. }
\end{figure*}

\subsection{The Detectability of Assembly Bias with Galactic Conformity}

Figure \ref{conformity_age} shows measurements of conformity for all
our various definitions of halo age. As outlines in the previous
subsection, primary galaxies are pure central galaxies and secondary
galaxies are are all central galaxies in the same stellar mass bin. In
both panels, we show our fiducial model using halo growth since
$z=0.8$ with the colored symbols. Error bars are obtained by jackknife
sampling based on RA and Dec. The left-hand side of Figure
\ref{conformity_age} shows the results for age-matching modeling in
which halos are ranked by fractional growth over various
timespans. For any definition, there is a clear conformity signal. In
fact, at small scales, $R_p\la 3$ \mpc, the conformity signal is
independent of redshift baseline. This is likely because halos that
have recently had a tidal interaction will appear as low-growth
halos in all models, but for short timescales, the halos that
interacted with each other will still be in proximity with one
another. At larger scales, there is a clear monotonic trend of a
larger conformity signal with longer redshift baseline, saturating for
$z\ga 0.8$. For $z=0.2\rightarrow 0$, there is almost no conformity
past $R_p=5$ \mpc.

In the right-hand panel of Figure \ref{conformity_age}, we show the
various formation epoch definitions. For $\zhalf$ and $a_c(\mhalo)$,
the conformity signal is roughly consistent and somewhat smaller than
the maximal effect seen for $z=0.8$ halo growth. However, when using
$a_c(\mpeak)$, the amplitude of the conformity signal is attenuated at
all scales, including small scales where recent tidal interactions
come into play. Because tidal interactions do not alter a halo's
$\mpeak$ value, the small scale two-halo conformity signal is much
smaller. 

\section{Results}

\subsection{Galactic conformity in $\fq$ for central galaxies}

Figure \ref{conformity_data} presents our measurements of $\fq$ around
pure central galaxies in the SDSS group catalogs, shown with the blue
and red symbols. Each panel shows results from a different stellar
mass bin. All are 0.2 dex wide. Recall that both primaries and
secondaries are restricted to the same stellar mass, and secondary
galaxies are centrals only. Error bars on the SDSS measurements are
obtained by dividing the sky area into 25 roughly-equal patches of sky
and performing jackknife sampling. The value of $\chi^2/\nu$ in each
panel is obtained comparing $\fqr$ and $\fqb$,

\begin{equation}
\chi^2/\nu = \frac{1}{N_{\rm data}} \sum
\frac{(\fqr-\fqb)^2}{\sigma^2_{\rm red} + \sigma^2_{\rm blue}},
\end{equation}

\noindent where $\nu=N_{\rm data}$ is the number of data points (9),
and $\sigma_{\rm red}$ and $\sigma_{\rm blue}$ represent the errors on
$\fqr$ and $\fqb$, respectively. For five of the six stellar mass
bins, there is no statistically significant evidence for a difference
between $\fqr$ and $\fqb$. The lone exception is the conformity
measurement for galaxies with $\log\mgal/\msol=[10.5,10.7]$, which
we will discuss subsequently.

In each panel, we show the predictions for two different age-matching
models: the $z=0.8$ model and the $a_c(\mpeak)$ model. At low stellar
masses, the differences between these two models is especially clear,
with $z=0.8$ producing a clearer signal, larger than that measured in
the data. At masses above $\mgal=10^{10.3}$ $\msol$, the predictions
of both models show only modest, if any, conformity.

Figure \ref{conformity_data} raises two pertinent questions: (1) What
are the $\chi^2$ values if we restrict our measurement to smaller
scales, where the conformity signal is predicted to be clearest, and
(2) If the conformity signal in the data were as strong as the
age-matching models, would we have been able to detect it given the
larger errors in the data? 

\begin{figure}
\vspace{-2.5cm}
\hspace{-1cm}
\includegraphics[width=7in ]{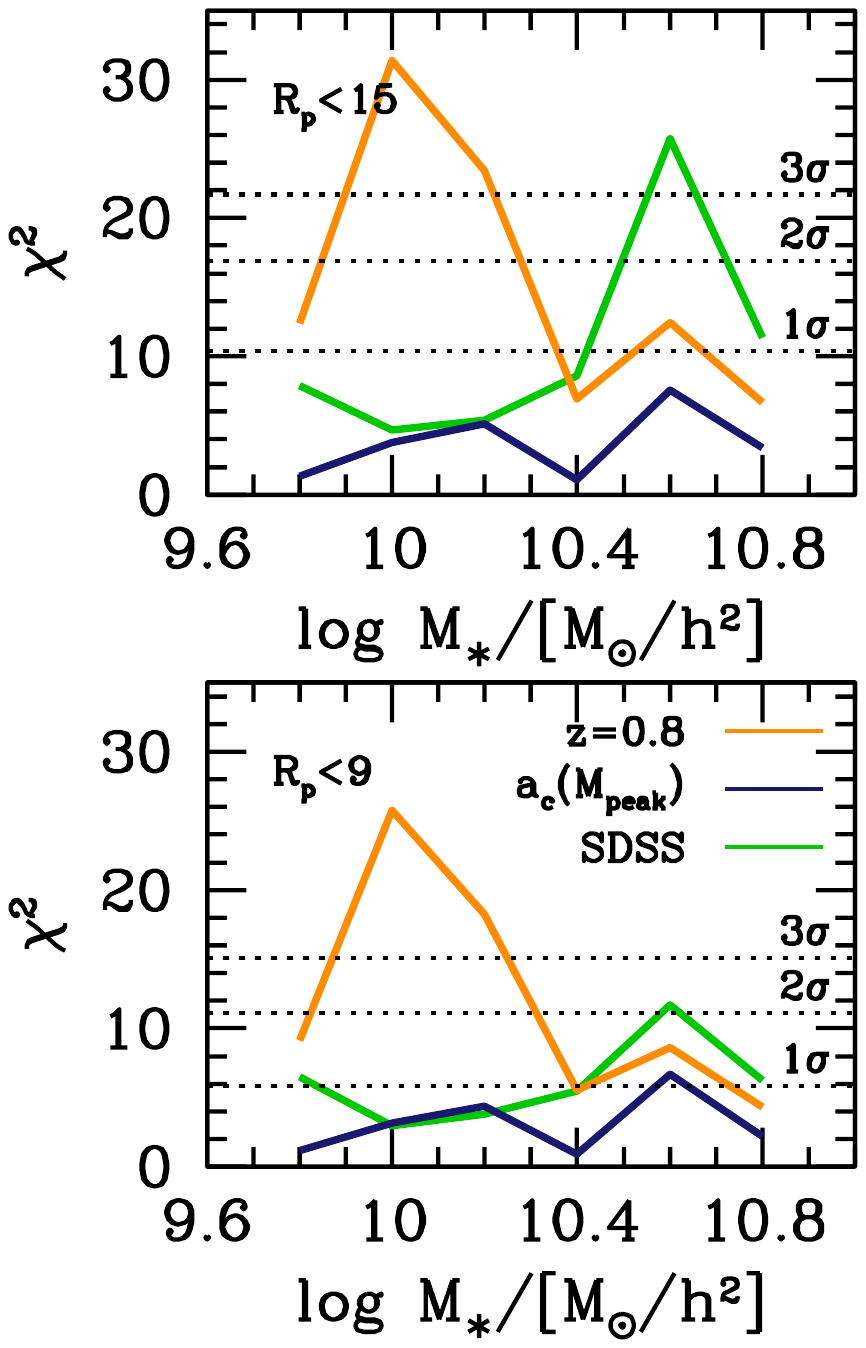}
\vspace{-8.0cm}
\caption{ \label{chi2} {\it Top Panel:} $\chi^2$ values, as defined by
Eq. (1), showing the difference in $\fqr$ and $\fqb$ as a  function of
$\mgal$. The three solid curves show results for the SDSS data, the
$z=0.8$ age-matching model, and the $a_c(\mpeak)$ model. The dotted
curves show confidence levels from a $\chi^2$ distribution for 9
degrees of freedom, representing the number of $R_p$ bins in the
measurement. For the age-matching models, the errors are taken from
the data, not from the mocks, which are significantly larger. Thus,
the curves represent the $\chi^2$ that would be obtained if the
model's conformity signal were present in the data. {\it Bottom
  Panel:} Same as the top panel, but now restricting all $\chi^2$
values to $R_p<6$ \hmpc\ bins (5 degrees of freedom). Note that the
$>3\sigma$ result from $\mgal=10^{10.3}$ $\msol$ is attenuates when
removing larger scales, implying that the result is partially driven
by a statistical fluctuation or that it is dependent on scale in a
manner not seen in any theoretical model. }
\end{figure}

Both of these questions are addressed in Figure \ref{chi2}.  In the
top panel, we show $\chi^2$ from Eq. (1)---without dividing by the
degrees of freedom---for the SDSS data as a function of stellar
mass. For reference, the $1\sigma$, $2\sigma$, and $3\sigma$
confidence levels from a $\chi^2$ distribution for 9 degrees of
freedom are shown with the horizontals dashed line. The other two
lines represent the $\chi^2$ values obtained from the $z=0.8$ and
$a_c(\mpeak)$ models after replacing the error bars on those
predictions with the error bars obtained from the SDSS group
catalogs. Thus, a $\chi^2$ above 16.5 indicates that, if the amount of
assembly bias seen in the age-matching model were present in the SDSS
data, it would be detectable at 95\% confidence. Unsurprisingly, for
the $a_c(\mpeak)$ age-matching model, the amount of conformity induced
is too weak to be detected at any stellar mass. The $z=0.8$ model
yields a $>3\sigma$ signal for lower stellar masses, where the
assembly bias in dark matter halos is strongest. The $\chi^2$ value
for the smallest stellar mass bin is barely above $1\sigma$, owing to
the small volume of this sample.

The lower panel of Figure \ref{chi2} shows the $\chi^2$ for the same
models and data, only now we exclude the data points at $R_p=9$ \mpc\
and above, reducing the number of degrees of freedom to 5. Relative to
the confidence levels, the results are consistent with those of the
full measurement. However, we note that the lone $>3\sigma$ detection
in the data, at $\mgal=10^{10.6}$ $\msol$, is now reduced to $\approx
2\sigma$. In all models of conformity, the signal is larger at smaller
separations. For the statistical significance to reduce when excluding
larger scales argues that this is a fluctuation, or due to some effect
that is distinct from galaxy assembly bias.

\begin{figure*}
\vspace{-2.8cm}
\includegraphics[width=5.2in, angle=90]{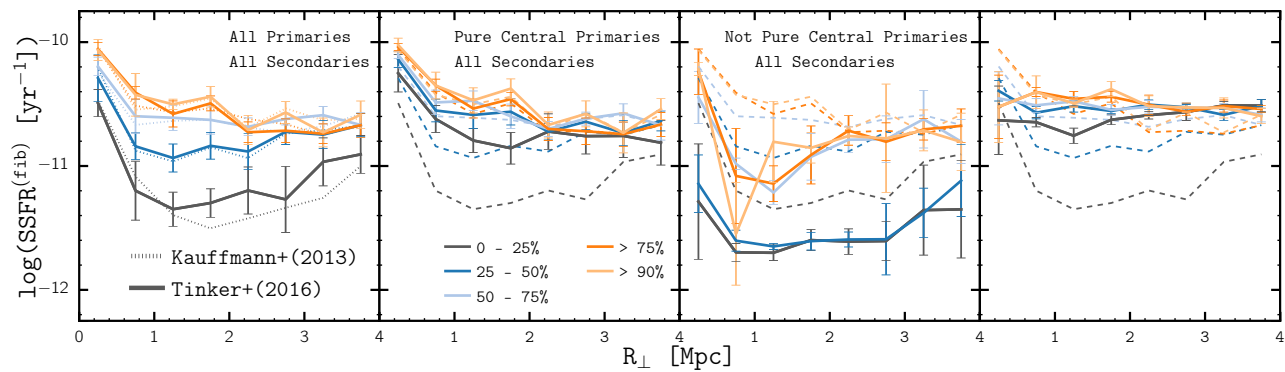}
\vspace{-4.2cm}
\caption{ \label{gk_conformity} Our
  reconstruction and subsequent deconstruction of the K13 conformity
  measurement. Panels from left to right: {\it Leftmost Panel:} The
  original K13 measurement: the median specific star formation rate
  within the SDSS fiber, \ssfr_fib, of secondary galaxies around
  primary galaxies isolated with the K13 criterion. Primary galaxies
  are in the stellar mass range $\log\mgal/\msol=[10.0,10.5]$, while
  secondary galaxies include all galaxies in the sample. The thin
  solid curves are taken from K13, and the thick curves with error
  bars are our own measurement. The error bars are obtained by spatial
  jackknife of the sample. {\it Second Panel:} Thick solid lines show
  the measurement after removing from the primary sample galaxies that
  are classified by the group finder as either satellite galaxies are
  non-pure satellites. This removes $\sim$6\% of galaxies from the
  primary sample. The dotted curves---here and in the other right-hand
  panels---show our measurement of conformity from the left-most
  panel. {\it Third Panel:} The conformity
  signal of the galaxies that were removed from the primary sample in
  the second panel. Half of these galaxies are classified as
  satellites, while the other half are classified as low-probability
  centrals. {\it Right Panel:} The conformity signal when restricting
  the secondary galaxies to be central galaxies of the same stellar
  mass as the primary sample. The volume of this catalog is larger
  than the other three panels, thus the error bars are smaller. }
\end{figure*}

\begin{figure*}
\includegraphics[width=6.6in]{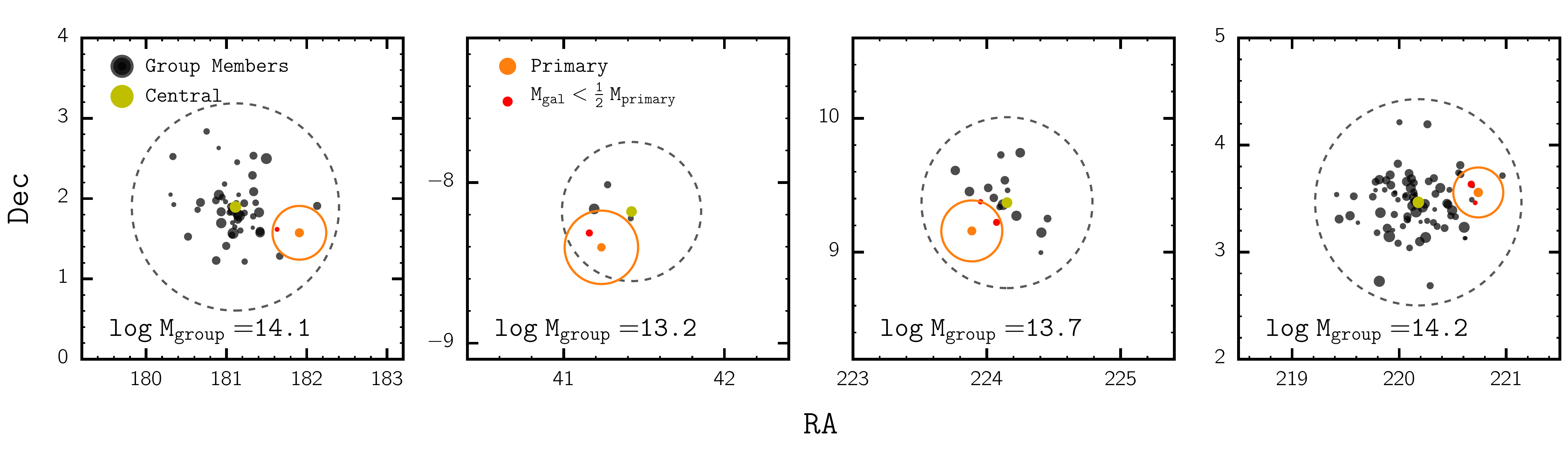}
\caption{ \label{gk_groups} Examples of galaxies classified as
  isolated by the K13 criteria, but are marked as satellites within a
  group by the group finder. In each panel, the isolated primary
  galaxy in question is marked in orange, with the isolation radius
  shown with the orange circle. The group virial radius is indicated
  with the dashed circle, while other group members are shown in gray,
  with point size proportional to $\log\mgal$. The group central
  galaxy is shown in green. Red dots indicate galaxies that are within
  the projected radius of the isolation criteria but are less massive
  than $\mgal/2$. Other group members (gray symbols) within the
  isolation radius are outside the velocity separation criterion
  ($\Delta v>500$ km/s).}
\end{figure*}

\subsection{Conformity in the Star Formation Rates of Galaxies}

How can our results be reconciled with the measurements of K13? They
are substantially different measurements, given our use of group
catalogs to identify centrals, and our restriction of the secondary
population to be centrals of the same stellar mass as the primaries,
the `quantity of conformity' being quenched fraction rather than
median specific star formation rate (sSFR). First, we reproduce the
K13 measurement, then explain the differences in our results. We will
demonstrate that our use of a group catalog to robustly identify
primary galaxies is the key difference in the comparison.

Figure \ref{gk_conformity} shows our reproduction of the K13
conformity measurement. For this measurement, all galaxy stellar mass
and star formation rates come from the MPA-JHU catalog as done in
K13. In the previous sections of this paper, we use NYU-VAGC stellar
masses.  K13 also uses a mixture of the total sSFR---corrected for the
finite aperture of the fiber relative to the angular size of the
galaxy---and the the sSFR only within the fiber aperture. We will
refer to the total sSFR as such and the rate within the fiber as
\ssfr_fib. We created a stellar mass complete sample incorporating all
galaxies with $0.017<z<0.030$ and $\mgal>10^{9.25}$ $\msol$. Primary
galaxies are identified within this sample using the isolation
criterion of K13: a galaxy with stellar mass $M$ is isolated if there
are no galaxies more massive than $M/2$ within a projected separation
of 500 kpc and a $\Delta v$ of 500 km/s. Secondary galaxies are all
galaxies within the sample. The primary galaxies are broken in
percentile bins based upon their total sSFR: 0-25\%, 25-50\%, 50-75\%,
$>$75\%, and $>$90\%. The $y$-axis represents the median \ssfr_fib\
for secondary galaxies around each bin in primary sSFR.  The left-hand
panel in Figure \ref{gk_conformity} compares our measurement to that
of K13: all primaries and all secondaries are used. Error bars are
from spatial jackknife subsampling of the SDSS footprint into 25
equal-area regions. There are slight differences in the median
\ssfr_fib\ for the lowest star-forming primaries, but the results are
consistent in general and, in particular, both measurements show a
strong conformity signal; neighboring galaxies around primaries with
suppressed star formation rates also show significantly lower specific
star formation rates. We find similar comparison to K13 when using
aperture corrected sSFR.

We have also run the group finder on this catalog, enabling us to
investigate the agreement between K13's isolation criterion and our
own, and to bring these measurements into a more common framework with
the conformity measurements of the quenched fraction earlier in the
paper. In the group catalog, 3.5\% of the K13 primary galaxies are
classified as satellites. Figure \ref{gk_groups} shows several
examples of galaxies that are classified as isolated according to K13
but are denoted as satellite galaxies in the group finder.
Additionally, another 3\% of the K13 primary galaxies are classified
as `non-pure' central galaxies (i.e., their $\psat$ values are $<~0.5$
but $>~0.01$). These two subpopulations represent only 6.5\% of the
K13 primary sample, but they have a dramatic impact on the measured
conformity signal. The second panel in Figure \ref{gk_conformity}
shows the conformity measurements for the 93.5\% of K13 primaries that
are also listed as pure centrals. The conformity signal is nearly
gone. The sSFR of secondary galaxies is roughly independent of the
sSFR of the primary galaxy, with some small differences at $1<R_p<2$
Mpc. Because the primary sample is made up of relatively massive
galaxies at $\log\mgal/\msol=[10.0,10.5]$, any satellites within this sample
will lie in a massive halo. These massive halos will, in turn, contain
a large number of quenched satellite galaxies up to $\sim 2$ Mpc away
from the satellite-primary galaxy (1 Mpc being the radius of
$\mhalo\sim 10^{14}$ $\msol$ halos). Although the overall fraction of
quenched galaxies is $\sim 50\%$ at $\log\mgal/\msol=[10.0,10.5]$, for
satellites at that mass scale, the quenched fraction is $\ga 80\%$,
thus the inclusion of these galaxies specifically biases the lowest
two bins in primary sSFR, as can be seen in Figure
\ref{gk_conformity}.

On the right-hand panel, we attempt an apples-to-apples comparison of
the conformity signals made on $\fq$ to sSFR. The primary galaxies
are, once again, pure central galaxies in the stellar mass range
$\log\mgal/\msol=[10.0,10.5]$, but now we restrict the secondary galaxies to
also be central galaxies within the same mass range. To enhance the
statistics in the measurement, we create a new stellar-mass limited
catalog for galaxies with $\mgal>10^{10.0}$ $\msol$ and
$0.017<z<0.0525$, and run the group finder on this catalog. When
constructed in the same manner as the $\fqcen$ measurements, the sSFR
conformity measurements are consistent.

The group finder is not infallible; from mock tests, roughly 20\% of
the central galaxies are actually misclassified satellites. However,
that number is falls to around 1.8\% overall, and 1.6\% in the mass
range for the K13 measurements, when restricting the sample to pure
centrals. Whether or not the labeling of a galaxy as `central' or
`satellites' in the group catalog is 100\% accurate, a conservative
interpretation of Figure \ref{gk_conformity} is that a more
restrictive isolation criterion essentially eliminates the conformity
signal seen in K13. Furthermore, any mislabeling of centrals and
satellites in our primary sample not eliminated by the purity cuts is
likely to {\it contribute} to a conformity signal, so the results in
the right-hand panel of Figure \ref{gk_conformity} are upper limits on
the true conformity.

\begin{figure}
\vspace{-2.5cm}
\hspace{-1cm}
\includegraphics[width=6.6in]{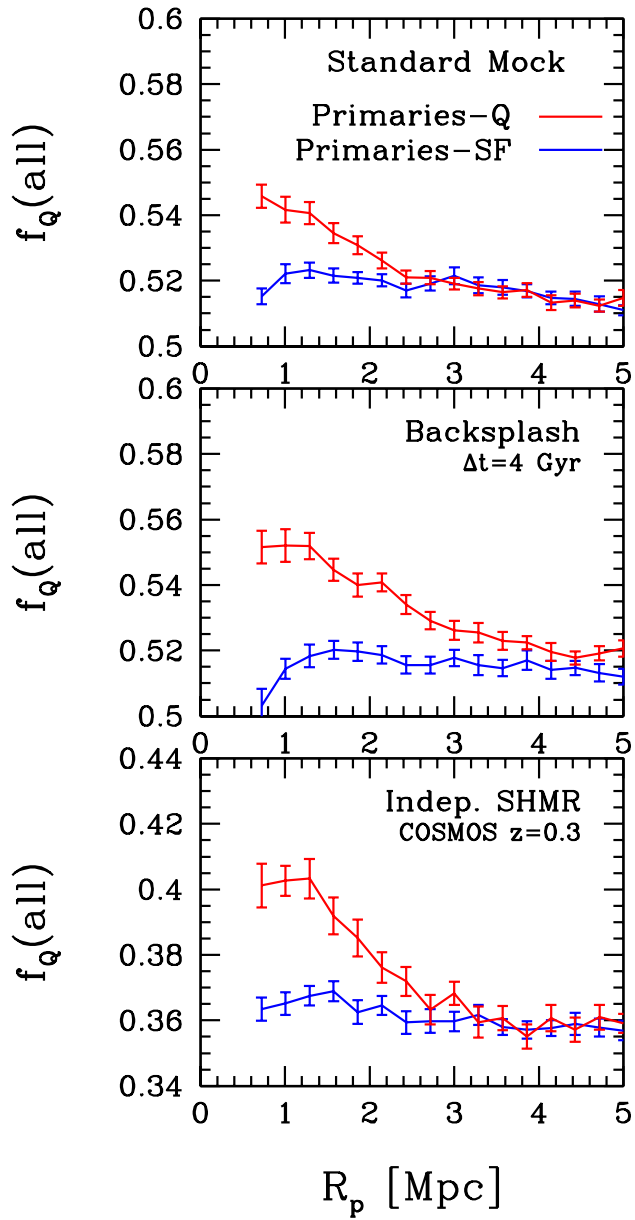}
\vspace{-2.5cm}
\caption{ \label{gk_tests} Three measurements of conformity using the
  K13 isolation criterion on mock galaxy samples. In each panel, the
  primaries at isolated galaxies in the stellar mass range
  $\log\mgal/\msol=[10.6,11.1]$, while the secondaries are all galaxies with
  stellar masses $\log\mgal/\msol>9.7$. Curves show the quenched fraction of
  secondaries around star-forming and quenched primaries. Error bars
  are from jackknife sampling. {\it Top Panel:} Mock with no assembly
  bias. This is the same mock used in Figure
  \ref{conformity_tests}. {\it Middle Panel:} Mock using the same
  galaxy catalog, but splashback galaxies with first accretion times
  more than 4 Gyr ago are classified as quenched. See text for more
  details. {\it Bottom Panel:} Mock where the stellar-to-halo mass
  relation of star-forming and quenched galaxies are taken from
  \citealt{tinker_etal:13_shmr}, in which the halo masses of quiescent
  galaxies are higher than those of star-forming galaxies at fixed
  stellar mass. The relations are calibrated to match statistics of
  galaxies in COSMOS data at $z=0.3$. See text for further details. In
  each panel, there is a marginal detection of conformity at $R_p<2$
  Mpc, even though each mock has no assembly bias.}
\end{figure}

\subsection{Conformity from alternate sources}

The results of the previous section indicate that the conformity
signal can be sensitive to the details of the isolation
criterion. Figure \ref{gk_tests} shows the conformity signal in
several different mock galaxy catalogs, none of which contain any
galaxy assembly bias (i.e., the values of $\dn$ do not correlate with halo
formation history). Here, conformity is measured as the quenched
fraction of secondary galaxies around star-forming and quenched
primary galaxies. Primaries are identified using the isolation
criterion of K13 on galaxies in the stellar mass range
$\log\mgal/\msol=[10.6,11.1]$. We choose this mass range because it yields a
median galaxy mass close to that recent conformity results from
\cite{berti_etal:16} using PRIMUS data, which also use the K13
isolation criteria. Secondary galaxies are all galaxies in the mock,
which is complete down to $\mgal=10^{9.7}$ $\msol$. We note that the
results shown in this figure are qualitatively the same when using a
sample of primaries in the range $\log\mgal/\msol=[10.0,10.5]$, as in K13.

The top panel in Figure \ref{gk_tests} shows the results for a mock
with no assembly bias. This is the same mock used in Figure
\ref{conformity_tests} to test our method of measuring conformity
using the group catalog. At $R_p>2$ Mpc, there is no
conformity. However, at smaller scales, there is a small but
measurable difference in the quenched fractions of secondary galaxies
around star-forming and quiescent primaries. At $R_p\sim 1$ Mpc, this
difference is around 2\%, driven mostly by the same effects seen in
the K13, in which a small fraction of satellite galaxies make it into
the primary sample.

In the middle panel, we incorporate the effects of backsplash galaxies
into the mock. As discussed above, backsplash galaxies are those that
are currently classified as central, but have in their past history
passed through the virial radius of a larger
halo. \cite{wetzel_etal:14} showed that the slight enhancement of the
$\fq$ around groups and clusters can be explained by a model in which
backsplash galaxies evolve the same as satellite galaxies: several Gyr
after the initial accretion event, the galaxies undergo rapid
quenching and migrate onto the red sequence. Most backsplash galaxies
are eventually re-accreted back into the larger halo, but some exist
as centrals long enough to be quenched while outside the group or
cluster's virial radius. Here, we identify all central galaxies that
are backsplash galaxies. If the initial accretion event took place
more than 4 Gyr ago, the galaxy is marked as quenched if it is not
already. This delay time is taken from the results of
\cite{wetzel_etal:14} to match the observed quenched fraction around
groups and clusters. In the mock, to offset the overall increase in
$\fqcen$, a random sample of quenched central galaxies are
reclassified as star forming in order to preserve the initial quenched
fraction. This process reclassifies about 5\% of the central galaxy
population. 

Backsplash galaxies are preferentially near large central
galaxies. The impact of this on the conformity signal is seen in the
middle panel of Figure \ref{gk_tests}. There is a slight difference in
the quenched fractions all the way out to 5 Mpc, caused by splashback
galaxies themselves being classified as primaries. But inside 2 Mpc
the difference in $\fq$ around star-forming and quenched primaries
monotonically rises to around 4\% at $R_p\sim 1$ Mpc,

The bottom panel shows the results of measuring conformity on a mock
constructed from the stellar to halo mass relations (SHMR) calibrated
on COSMOS data in \cite{tinker_etal:13_shmr}. The key difference
between this mock and the ones used elsewhere in this paper is that
\cite{tinker_etal:13_shmr} use stellar mass functions, clustering, and
galaxy-galaxy lensing measured separately for passive and star-forming
galaxies to constrain the relationships between halo mass and galaxy
mass independently for the two classes of galaxies. In works like
\cite{hearin_watson:13}, it is assumed that passive and star-forming
galaxies live of the stellar mass live in halos of the same dark
matter mass. There is no reason {\it a priori} that this should be
true, and the results of \cite{tinker_etal:13_shmr} show significant
differences between the halos of red and blue galaxies, especially at
high stellar mass (see also, \citealt{more_etal:11, mandelbaum_etal:16,
  zu_mandelbaum:16}). Massive quiescent galaxies preferentially occupy
more massive halos than star-forming galaxies of the same mass at
$z=0.3$.

To create this mock, we use the $z=0.3$ output of the MultiDark
Planck-2 simulation (MDPL2; \citealt{prada_etal:13}), which is
publicly available for download\footnote{\tt
  https://www.cosmosim.org/cms/simulations/mdpl2/}. Subhalos are
discarded and only host halos are used. Host halos are populated using
the SHMRs for passive and active galaxies, including central and
satellite populations. The cosmology assumed in
\cite{tinker_etal:13_shmr} has $\om=0.27$, while MDPL2 has
$\om=0.306$. To correct for this, we increase the halo mass scales in
the SHMR fitting functions by a factor of $0.306/0.27=1.13$, which
mostly corrects for the change in cosmology. As shown in
\cite{leauthaud_etal:12_shmr}, the COSMOS stellar masses are roughly
0.2 dex larger than the {\tt kcorrect} stellar masses used in the
VAGC, so we shift all stellar masses down by 0.2 dex to put them on
the same scale as the other mocks. 

The results are shown in the bottom panel of Figure
\ref{gk_tests}. Because passive galaxies live in more massive halos
than star-forming galaxies of the same stellar mass, the environments
probed by passive primary galaxies differs from that of star-forming
primary galaxies. At $R_p<3$ Mpc, $\fq$ around passive primaries shows
a increase over star-forming primaries, increasing to around $\sim
4\%$ at $R_p=1$ Mpc.

\section{Discussion}

\subsection{Comparison to Previous Work: Theoretical}

We have shown that an the age-matching model can indeed produce
strong galactic conformity out to $\sim 10$ Mpc. In fact, most
standard age-matching models that produce a strong signal at small
scales ($\sim 2$ Mpc) produce a signal at large scales. These results
are in qualitative agreement with the previous results in
\cite{hearin_etal:15}. However, the amplitude of the conformity signal
does depend on how age is defined. Models in which halo age is defined
using peak halo mass rather than current halo mass---a process which
limits the impact of tidal encounters on halo age---produces a much
smaller conformity signal at all scales. 

\cite{bray_etal:16} analyze galaxies in the Illustris cosmological
hydrodynamic simulation in order to determine the conformity from that
model of galaxy formation. \cite{bray_etal:16} find that the
conformity signal in galaxies is roughly the same as the conformity
signal in halos (after dividing the halo population up into old and
young subsets). The amplitude of the conformity, once restricted to
primary and secondary galaxies both being central, is consistent with
what we find in our standard age-matching models. Our $a_c(\mpeak)$
model yields a smaller conformity signal than found in Illustris, thus
we conclude that tidal encounters have an immediate impact on star
formation in galaxies in Illustris. 

We demonstrated that backsplash galaxies can produce a conformity
signal of a few percent at $R_p\la 3$ Mpc. This is in contrast to
\cite{hearin_etal:15}, who analyzed a backsplash model based on the
results of \cite{wetzel_etal:14} . They found no statistical evidence
for conformity produced by such models. In this paper, we have used a
simulation with four times the volume, increasing the statistical
precision of the model and revealing the conformity signal seen in
Figure \ref{gk_tests}. Additionally, we have implemented the K13
isolation criterion to fully incorporate any observational effects.

\subsection{Comparison to Previous Work: Observational}

For large-scale (a.k.a. `two-halo') conformity, there is a surprising
dearth of measurements for the local universe. The K13 measurements
are specifically about specific star formation rates of secondary
galaxies around primary galaxies. This work represents the first
measurement of conformity using $\fqcen$ as the statistic of
interest. After correcting for contamination in the isolation
criterion of K13, and restricting the secondary galaxies to be
centrals of the same stellar mass range as the primaries, the
conformity signal in sSFR is much closer to consistent with the
$\fqcen$ results here.


\cite{hatfield_jarvis:16} measured the angular clustering of
photometrically selected galaxies. At $z\sim 1$, they cross-correlated
high-mass and low mass galaxies, breaking both samples into passive
and star-forming objects. They found that the amplitude of the
cross-correlation function of high-mass with low-mass passive galaxies
is higher than high-mass galaxies crossed with low-mass star-forming
galaxies. This is consistent with the effects of assembly bias, but it
is difficult to disentangle the degeneracies in clustering amplitude
between assembly bias, satellite fractions of red and blue galaxies,
and the fact that red and blue galaxies of the same mass may occupy
different mass halos.

\cite{berti_etal:16} also use PRIMUS data to probe conformity in the
redshift range $0.2<z<1$. In \cite{berti_etal:16}, conformity is quantified by
finding isolated `central' galaxies using the K13 isolation criterion,
and then measuring $\fq$ around passive and star forming primaries. At
$R_p<2$ Mpc, they find that $\fq$ around passive primaries is between
1-4\% higher than around star-forming primaries. The signal detected
is statistically robust, but is consistent with the amount of
conformity seen in \S 3.3, either from backsplash galaxies or
different halo occupation for star-forming and passive
galaxies. \cite{berti_etal:16} finds that $\fq$ around the two samples
is consistent at $R>2$ Mpc, which is inconsistent with the predictions
of standard age-matching; in Figure \ref{conformity_age}, standard
age-matching models show a signal out to 10 Mpc and beyond. The
\cite{berti_etal:16} results are consistent with the expected
age-matching signal from the $a_c(\mpeak)$ model as well as a model in
which recent halo growth is used to define halo `age'. But using
conformity alone it is impossible to distinguish the assembly bias
effect in this model from the effects of splashback galaxies and
differential halo occupation.

\subsection{One-Halo and Two-Halo Conformity}

Although the observational picture of large-scale conformity is murky
at best, there is broad consensus that small-scale conformity---in
which satellites within a dark matter halo are more likely to be
passive if the central galaxy within that halo is passive---exists in
the galaxy distribution. A number of authors, using different methods,
have confirmed the original measurement of \cite{weinmann_etal:06}
(\citealt{phillips_etal:14, knobel_etal:15, kawinwanichakij_etal:16,
  berti_etal:16}).  

One explanation for small-scale conformity is indelibly tied to halo
assembly bias; older halos are more likely to have subhalos that are
older---i.e., they were accreted longer ago---than younger halos that
are growing more rapidly. Older subhalos are more likely to be
quenched of their star formation (\citealt{weinmann_etal:10, wetzel_etal:13}),
thus a correlation between host halo age and satellite galaxy colors
fits naturally in this model. However, in order to produce conformity,
the central galaxies in older halos must be more likely to be quenched
as well; i.e., two-halo conformity must exist because older halos will
be clustered with one another. How is it possible to
achieve one-halo conformity without such conformity existing at larger
scales?


\cite{kauffmann:15} proposes that AGN heating---a mechanism widely invoked to quench star formation in central galaxies---can heat halo gas, causing increased efficiency of quenching nearby galaxies through ram pressure or reduced gas accretion. This mechanism is proposed to explain the K13 conformity signal that reached out to large scales, but should also apply to scales within the virial radius.

One caveat on the detection of small-scale conformity is that the halo
masses are inferred or assumed, and biases may exist. In the group
catalog approach of \cite{weinmann_etal:06}, halo masses are assigned
using the total stellar mass of the group. For $\mhalo\la 10^{13}$
$\msol$, the total mass is dominated by the central galaxy. Thus, the
group finder assumes near one-to-one correlation between central
galaxy stellar mass and host halo mass, regardless of whether the
galaxy is star forming or quiescent. As discussed in \S 3.3, this
assumption that star-forming and quiescent galaxies of the same
$\mgal$ live in halos of the same dark matter mass is not supported by
studies constraining the halo occupation of these two types of
galaxies independently. Thus halo masses assigned in group catalogs
may be biased when split on central galaxy color: the halos
around quiescent central galaxies would be underestimated, while those around
star forming central galaxies would be overestimated. This could
impart some conformity between the properties of satellite galaxies and their host centrals when using group-catalog halo masses, but the impact of this bias has not been quantified. We leave more thorough investigation of this effect to another paper in this series. Within
semi-analytic galaxy formation models, \cite{wang_white:12}
found that massive, isolated, quiescent galaxies have more red
satellites than their star-forming counterparts for this very reason:
they live in more massive halos.

\subsection{Halo Formation and Central Galaxy Quenching}

Isn't the scenario described above---in which star-forming and passive
galaxies of the same stellar mass occupying different halos---itself a
manifestation of galaxy assembly bias? Possibly, but a correlation between galaxy assembly and halo assembly is not
required to create this scenario. If the process by which galaxies
stop forming stars is entirely stochastic at fixed halo mass, but the
efficiency of this process increases monotonically with $\mhalo$, you
naturally end up with passive galaxies living in higher mass halos at
fixed stellar mass: the halos of both passive and star forming
galaxies continue to grow, but only the star forming galaxies increase
their mass. Thus, at fixed halo mass, the passive galaxies are smaller
than the star forming ones (which translates into higher halo masses
for passive galaxies at fixed halo mass).

However, in order to wipe out any correlations with halo assembly
history in the present-day universe, the efficiency of the quenching
mechanism can only be correlated with the eventual $z=0$ halo mass,
and not the mass of the halo at the time at which the quenching
occurs. For example, if quenching occurs at some threshold in halo
mass, early-forming halos will reach that threshold earlier than
later-forming halos. The same correlation would exist if the threshold
for quenching was in galaxy stellar mass, under the assumption that
early-forming halos would have more massive galaxies even if all halos
converted the same fraction of accreted baryons into stars (see, e.g.,
\citealt{tinker:16}). Even a model in which quenching is not due to
crossing a threshold but rather due to a process which has imparts a
quenching probability that varies continuously---and probably
monotonically---with halo mass and redshift would contain some imprint
of halo formation history because the early-forming halos would have a
higher quenching probability at a given redshift. The degree to which
this is represented in the spatial distribution of quenched galaxies
will in large degree be reflective of how steep the quenching
probability is with $\mhalo$.

From Paper I and this paper, we have observational evidence that the
correlation between the quenching mechanism and halo formation history
is weak but non-zero for massive galaxies (shown in Paper I) and close
to negligible for lower mass galaxies (shown in both papers). This
supports a model in which quenching of galaxies is a stochastic
process, where the probability of going through the quenching process
is a weak function of mass at low masses and strong function of
mass---closer to a threshold---for higher masses. The evidence
presented here does not indicate {\it which} mass is most important:
whether the quenching probability is determined by $\mhalo$ or
$\mgal$. Because of the tight correlation between the two, a quenching
threshold in one property would induce a correlation between quenching
probability and the other quantity. \cite{tinker:16} proposes using
the scatter in the relationship between stellar mass and halo mass to
distinguish between these two scenarios. \cite{gu_etal:16} have demonstrated that star formation, rather than merging, is the dominant contributor to the this scatter for all but the most massive galaxies and halos.

With the measurements put forward in this paper and Paper I, in
combination with other measurements, there should now be enough data
to constrain the relationship between central galaxy quenching and
halo mass and formation history. This combination includes the scatter
in relationship between halo mass and stellar mass, the fraction of
central galaxies that are quenched and how this quantity depends on
$\mgal$ and redshift, constraints on the SHMRs of star-forming and
quiescent galaxies, and the existence of small-scale galactic
conformity. These data present a wealth of information to constrain
how the quenching probability depends on $\mhalo$ and redshift. Armed
with this knowledge, we will take a major step forward in
understanding which physical mechanisms are most important for
quenching, which are secondary correlations without causation, and
which are uncorrelated with the process that stops stars from being
formed in central galaxies.


\section{Conclusions}

We have measured the galactic conformity signal around central galaxies using group catalogs to isolate `primary' central galaxies. The quantity we measure is the quenched fraction of central galaxies around primary centrals galaxies that have been divided up by their star formation activity---i.e., a quenched sample and a star-forming sample.
We then compare these measurements to different theoretical models that vary in how galaxy stellar age is correlated (or not) with halo age, using various definitions of halo ago. We focus on scales larger than 1 Mpc in order to isolate effects between pairs of distinct halos, rather than galaxies that share the same host halo. We find the following:

\begin{itemize}
\item In SDSS DR7 data, there is little to no statistical evidence of any difference in the $\fq$ of central galaxies around star-forming centrals and quiescent centrals. 
\item If galaxy quenching were correlated with halo age, using halo age definitions such as $\zhalf$, $a_c(\mhalo)$, and halo growth since $z=0.8$, there would be a significantly detectable ($\ga 3\sigma$) signal of galactic conformity. We do not detect this.
\item Other definitions of halo age, such as $a_c(\mpeak)$ and halo growth over shorter redshift intervals like $\Delta z=0.1$, would not yield a detectable conformity signal in the data.
\item The strong conformity signal in galaxy star formation rates seen in K13 is almost entirely eliminated by removing a small number of satellite galaxies that are not excluded in the K13 isolation criterion.
\item At $1<R<3$ Mpc, small conformity signals in $\fq$ can be created by means other than galaxy assembly bias. 
\end{itemize}


\vspace{1cm}
\noindent The authors wish to thank Michael Blanton and Risa Wechsler
for many useful discussions. 
A.R.W. was supported by a Caltech-Carnegie Fellowship, in part through the Moore Center for Theoretical Cosmology and Physics at Caltech. The authors thank Matthew R. Becker for
providing the Chinchilla simulation used in this work. The Chinchilla
simulation and related analysis were performed using computational
resources at SLAC. We thank the SLAC computational team for their
consistent support.
The authors acknowledge the
Gauss Centre for Supercomputing e.V. (www.gauss-centre.eu) and the
Partnership for Advanced Supercomputing in Europe (PRACE,
www.prace-ri.eu) for funding the MultiDark simulation project by
providing computing time on the GCS Supercomputer SuperMUC at Leibniz
Supercomputing Centre (LRZ, www.lrz.de).  The Bolshoi simulations have
been performed within the Bolshoi project of the University of
California High-Performance AstroComputing Center (UC-HiPACC) and were
run at the NASA Ames Research Center.


\bibliography{../risa}

\label{lastpage}

\end{document}